\documentclass[12pt]{article}
\usepackage[a4paper,margin=1in]{geometry}
\usepackage{amsmath,amssymb,bm}
\usepackage{siunitx}
\usepackage{booktabs}
\usepackage{graphicx}
\usepackage{caption,subcaption}
\usepackage{hyperref}
\usepackage{setspace}
\title{\bf {From Frequency Dependent Specific Heat to Fictive Temperature of a Glassy Liquid}}
\author{\bf Biman Bagchi, \\ SSCU, Indian Institute of Science, Bengaluru 560012, INDIA}
\date{\today}
\begin{document}
\maketitle
\begin{abstract}
Upon rapid quenching of temperature of a glass forming liquid, the system falls out of equilibrium due its finite relaxation time. Additionally, the relaxation becomes progressively slower with time. The created nonequilibrium state of the glassy system is conveniently described by introducing a fictive temperature which provides the instantaneous state of the nonequilibrium system. The fictive temperature $T_{f} (t)$  is time dependent. During cooling, the fictive temperature is higher than the actual temperature. After the cooling or quenching has ceased, the fictive temperature approaches the final temperature at a rate that depends on the relaxation properties of the liquid.  In this work we use linear response theory to connect the time dependence  of the fictive temperature to memory function which is shown to be related to the frequency dependent specific heat which itself depends on the fictive temperature $T_{f} (t)$. Thus, one requires { \it a self-consistent calculation} to capture the interdependence of relaxation rate and structural response function. We present a numerical calculation where we apply our relations to  silica where the relaxation function that describes the frequency dependent specific heat and is modeled as a stretched exponential William-Watts (WW) function, while the relaxation time is modeled as a Vogel-Fulcher-Tammann (VFT). We calculate the fictive temperature self-consistently. $T_{f}(t)$ exhibits the fall out from actual temperature as time (t) progresses.  
\end{abstract}
\section{Introduction}
Theoretical description of time evolution of a newly created non-equilibrium state of any system is a subject of much current discussion, and is not well-understood. Relaxation of a glassy liquid provides an example where, despite much efforts, success has been slow to come.
Non-exponential relaxation is a hallmark of supercooled liquids. Experiments often measure the time dependent response of such liquids to small perturbations. the relaxation back to equilibrium is found to be non-exponential and one employs non-exponential functions like Kohlrausch-Williams-Watts (KWW) stretched exponential form.\cite{KWW} However, when the liquid is subjected to a large perturbation, like a sudden quench of temperature, the resulting relaxation usually cannot be described by such closed form expressions. The response is also non-linear, and depends on the magnitude of the change. That is, the relaxation may not be described by using the conventional linear response theory. The objective of this work is to address this rather fundamental aspect of relaxation in systems far from equilibrium.

In our theoretical study,  a glass forming liquid is cooled at such a rate that the relaxation of the liquid cannot adjust to the rapidly falling temperature, the liquid falls out of equilibrium. That is, the structure and dynamics of the liquid remain trapped at values appropriate for higher temperatures. It is thus not possible to describe the state of the system by using the actual temperature which we denote by $T(t)$.There are many consequences of this fall out which have been subjects of study over many years, starting with the pioneering works of Tool many decades ago. [2] It was subsequently developed by Narayanaswamy and Moynihan [3,4]. It has been studied by Nielsen and Dyre [5] and by strunk [6].\\
 The first point to note is that as the liquid relaxes in its non-equilibrium state toward the glass transition, the viscosity grows strongly which slows down the relaxation. The system retains the memory of the past.
 
 Second, the system, due to its increasingly slow relaxation, lags behind the relaxation rate at the actual temperature. That is, the system behaves as if it is trapped at a higher temperature (during cooling). As the temperature gets lower, the relaxation becomes slower. The effect is non-linear meaning that the relaxation depends not only on the amplitude of the temperature change, but also on the initial and final temperatures, not just the difference. Thus, the system retains the memory of the initial state.
 
 In a landmark paper, Tool introduced the concept of fictive temperature to describe the system falling out of equilibrium. The \emph{fictive temperature} $T_f(t)$ is introduced to describe structural and dynamical properties which is reflective of a higher temperature.\cite{Tool1946} The fictive temperature at a given time can be significantly larger than the actual temperature $T(t)$. Both the temperatures, the actual ($T(t)$) and the fictive,  are time dependent. They are identical at time t=0 and also become identical in the long time. But at the intermediate times, the fictive temperature is higher than the actual time varying temperature. It is a difficult problem to determine $T_{f}(t)$.

 The time dependence of fictive temperature $T_{f}(t)$ is a useful theoretical quantity to describe the progression of the non-equilibrium state created by rapid quench of temperature. The fast degrees of freedom can adjust to the actual, changing temperature $T(t)$, except for infinitely fast cooling, but the slow degrees of freedom like the atomic and the molecular rearrangements, proceed at much slower pace.
\subsection{Kinetic models, fictive temperature, and connection to Ising--Glauber dynamics}

The present fictive-temperature formulation is closely related in spirit to earlier kinetic work from Bagchi group on glassy heat capacity and aging. In the model of Chakrabarti and Bagchi\cite{ChakrabartiBagchi2004}, the glass-forming liquid is represented by a mesoscopic collection of $\beta$ processes, each described as a thermally activated \emph{two-level system}, and an $\alpha$ process described as a cooperative transition in a double-well that is mediated by the $\beta$ relaxations. Within this framework, they computed the heat capacity $C_p(T)$ on a cooling–heating cycle and showed that the experimentally observed overshoot during heating arises largely from \emph{delayed energy relaxation} in the cooperative double-well. The same kinetic model yields scan-rate dependent glass-transition temperatures $T_g(\dot T)$ and a limiting fictive temperature $T_f^{\mathrm{L}}$ in qualitative agreement with classical Tool–Narayanaswamy analysis, thereby providing a microscopic realization of the fictive temperature concept.\cite{Tool1946},  \cite{Narayanaswamy1971} In a follow-up paper, Chakrabarti and Bagchi extended this landscape-based model to compute the \emph{frequency-dependent} heat capacity $C_p^\ast(\omega)$ and demonstrated a two-step ($\alpha$+$\beta$) spectrum whose low-frequency peak position follows a non-Arrhenius temperature dependence, in line with both experiments and simulations on supercooled liquids.\cite{ChakrabartiBagchi2005}

In another, previous study, Bagchi employed an Ising model with Glauber dynamics, to describe the time dependence of the fictive temperature. \cite{Bagchi1989GlassBook}

Although these works did not explicitly invoke a lattice Ising Hamiltonian, the underlying structure is very close to a kinetic Ising description: each local two-level system can be mapped to a spin variable $s_i=\pm 1$, cooperative $\alpha$ events correspond to concerted flips in a small cluster, and the stochastic hopping kinetics define a master equation with local detailed balance. In this sense the Chakrabarti–Bagchi fictive-temperature model already realizes an \emph{effective} kinetic Ising picture in the energy-landscape space, where the fictive temperature $T_f(t)$ is defined via the configurational enthalpy $H_c(t)$ and its evolution is governed by a Markovian rate matrix. A more explicit bridge to genuine Ising–Glauber dynamics comes from later work by Biswas and Bagchi,\cite{BiswasBagchi2010} who introduced a finite-length kinetic Ising chain (with spins coupled at the boundaries to mimic surface pinning) and solved the Glauber master equations to study the propagation and annihilation of dynamical correlations in confined fluids. Using single-spin-flip Glauber dynamics,\cite{Glauber1963} they showed how fast and slow relaxation channels, non-exponential decay, and effective homogeneous dynamics emerge from a simple Ising model. Taken together, these studies provide both (i) a kinetic landscape model that naturally defines fictive temperature and its scan-rate dependence, and (ii) an explicit Ising–Glauber framework within the same group. The present work can thus be viewed as combining these strands: the Tool/TNM fictive temperature $T_f(t)$ is placed on a firm kinetic footing, while an explicit Ising–Glauber dynamics offers a minimal lattice implementation of the same concepts, with the two-level systems of the earlier model now realized as spins evolving under Glauber rates.

\subsection {Frequency-dependent specific heat and enthalpy fluctuations }
A direct microscopic connection between enthalpy–enthalpy time-correlation functions and the
complex, frequency-dependent specific heat $C_p^{*}(\omega)$ was later developed by Saito,
Ohmine, and Bagchi.\cite{SaitoOhmineBagchi2013} 
In this work, the authors computed the equilibrium energy-fluctuation spectrum of supercooled
water from extensive molecular dynamics simulations and showed that $C_p^{*}(\omega)$ exhibits
a clear two-step ($\beta$+$\alpha$) structure, with a slow structural contribution emerging
at low frequencies and a fast vibrational component dominating at high frequencies.  
The analysis demonstrated that the imaginary part $C_p''(\omega)$ is directly proportional to
the Fourier transform of the enthalpy–enthalpy correlation function, thereby placing the
frequency-dependent specific heat 
on the same dynamical footing as dielectric and mechanical susceptibilities.
This work is particularly relevant in the present context because it provides an explicit,
simulation-based validation of the fluctuation–response identity used here to express the Tool structural kernel $\Phi(t)$ in terms of $C_p^(\omega)$.
When combined with the Chakrabarti–Bagchi kinetic model for glassy enthalpy relaxation,
the Saito--Ohmine--Bagchi results supply a deeper microscopic foundation for linking the
nonequilibrium fictive temperature $T_f(t)$ to equilibrium thermal fluctuations of the liquid.

Our objectives are many fold.  (i) First, we aim to present a transparent \emph{memory--kernel} formulation of the Tool relation for $T_f(t)$, (ii) second, we relate the structural response function to the frequency dependent specific heat, and (iii) to connect the time scale to a Maxwell viscoelastic model using a VFT viscosity for fused silica.
%
\section{Theoretical analysis : Linear response theory}

Let us consider a specific example, that of cooling of  fused silica linearly from the melt. The relevant parameters are presented later.

A measure of the structural relaxation time can be obtained from the Maxwell viscoelstic relaxation time $\tau_{s} = \eta_{0}/G_{\infty}$. We shall further discuss the values of relevant parameters as we progress.

The shear viscosity increases rapidly as the temperature is lowered. We assume it to be given by Vogel-Fulcher-Tammann (VFT) form. This makes the relaxation time to go to infinity at the glass transition temperature which is unrealistic for the slow aging of glass. This itself is an interesting issue. here we have capped the relaxation time at a very high value so that calculations can be carried out at temperatures below the glass transition temperature. This aspect requires further work.

%
%
\subsection{Linear Response Based Analysis}

The evolution of the fictive temperature can be formulated consistently within the linear–response and memory–function frameworks. the apprroach, although approximate, provides a good starting point in providing a microscopic understanding of the phenomena. What is more important, it introduces the memory function using the linear response theory with its time correlation function formalism, and leads to a connection with frequency dependent specific heat. the latter is defined as an equilibrium quantity as the fictive temperature. This of course lies at the heart of the fictive temperature logic.

At constant pressure, the configurational enthalpy $H(t)$ of a non-equilibrium liquid
relaxes toward its equilibrium value corresponding to the instantaneous bath temperature $T(t)$,
but the relaxation rate at any instant depends on the current structural state, quantified by the fictive temperature $T_f(t)$.
This leads to a generalized energy–balance equation of the form
\begin{equation}
\frac{dH(t)}{dt}
=-\!\int_0^t K(t-t';T_f)\,
\big[\,H(t')-H_{\mathrm{eq}}(T(t'))\,\big]\,dt',
\label{eq:Hbalance}
\end{equation}
where $H_{\mathrm{eq}}(T)$ is the equilibrium enthalpy the system would have if it were instantly equilibrated at temperature $T$,
and $K(t;T_f)$ is a memory kernel that quantifies how past deviations of the enthalpy from its instantaneous equilibrium value
relax within the structural environment characterized by $T_f$.
%
Equation~(\ref{eq:Hbalance}) asserts that the rate of enthalpy change at time $t$ depends on
(i)~how far the system is out of equilibrium at each earlier time $t'$,
measured by the difference $H(t')-H_{\mathrm{eq}}(T(t'))$, and
(ii)~how efficiently the structure present at $t'$ (with fictive temperature $T_f(t')$)
can dissipate that excess energy, represented by the kernel $K(t-t';T_f(t'))$.
In equilibrium, $T_f=T$ and the kernel reduces to a stationary function $K_{\mathrm{eq}}(t-t')$.
\subsection {Relation to microscopic fluctuations.}

We noe use linear response theory of time dependent statistical mechanics to express the memory function in terms of time correlation function. 
The obvious dynamical variable of interest is the enthalpy which is time dependent whose relaxation can be described by a memory function equation. The memory function itself can be related to enthalpy-enthalpy time correlation function which then paves the way to establishing a connection with the frequency dependent specific heat.
Close to equilibrium, one may evaluate $K(t;T_f)$ from the equilibrium enthalpy–enthalpy
time‐correlation function computed at the same structural state (approximated by $T_f$):
\begin{equation}
K(t;T_f)\;\simeq\;
\frac{1}{k_B T_f^2}\,
\frac{d}{dt}\langle \delta H(t)\,\delta H(0)\rangle_{T_f}.
\label{eq:Kdef}
\end{equation}
This expression follows from the fluctuation–dissipation theorem:
the memory kernel governing relaxation is proportional to the time derivative of the equilibrium
enthalpy correlation function evaluated for a structure equilibrated at $T_f$.
In nonequilibrium conditions, this approximation corresponds to assuming that
the structure evolves slowly enough that, over the duration of a single relaxation event,
the system behaves as if it were equilibrated at its instantaneous fictive temperature.

Fourier transforming and invoking the fluctuation–dissipation theorem gives
\[
K(\omega;T_f)\;=\;\frac{i\omega\,\Delta C_p(T_f)}{C_p^{\ast}(\omega;T_f)-C_p^{\infty}},
\]
where $C_p^{\ast}(\omega;T_f)$ is the complex specific heat evaluated for a structure equilibrated at $T_f$.
Insertion into the enthalpy balance and conversion from enthalpy to temperature through 
$H(t)-H_{\mathrm{eq}}(T)\!=\!C_p^{\mathrm{conf}}[T_f(t)-T(t)]$ yields the generalized Tool equation,

\begin{equation}
\dot T_f(t)= -\int_0^t M(t-t';T_f)\,[T(t')-T_f(t')]\;dt',
\qquad
M(\omega)=\frac{i\omega\,\Delta C_p/C_p^{\mathrm{conf}}}{C_p^{\ast}(\omega)-C_p^{\infty}} .
\label{eq:ToolMemory}
\end{equation}

Here we have changed the notation to memory function $M(t,T_{f})$ for convenience.
A rather detailed derivation of Eq.3 is given in the Appendix.
%
%
\subsection {Connection with the structural relaxation function.}

We now make a further transformation, to establish connection to equilibrium which is employed to describe functions at the fictive temperature.
At equilibrium the configurational contribution to the enthalpy relaxation is governed by a normalized structural relaxation function 
$\Phi(t)$:
\[
\frac{dH_c(t)}{dt}=-\frac{1}{\tau}\!\int_0^t\! \Phi(t-t')\,\frac{dH_c(t')}{dt'}\,dt',
\qquad
\Phi(0)=1,\;\Phi(\infty)=0.
\]
Transforming to frequency domain using Eq.3 gives 
$C_p^{\ast}(\omega)=C_\infty+\Delta C_p\,[1-i\omega\widehat{\Phi}(i\omega)]$, so that
\[
\frac{\delta T_f(\omega)}{\delta T(\omega)}=\frac{C_p^{\ast}(\omega)-C_\infty}{\Delta C_p}
=1-i\omega\,\widehat{\Phi}(i\omega).
\]
Inverse transformation of the above equation yields the Volterra-type integral equation in the time domain ‐ for $T_f(t)$,
\begin{equation}
T_f(t)
=T(t)-\!\int_0^t\!\Phi(t-s)\,\dot T(s)\,ds ,
\label{eq:ToolTime}
\end{equation}
which is the canonical linear‐response form of Tool’s equation.
%
\subsection {Reduced time : The internal clock}
%
The assumption of fictive temperature is equivalent to the assumption of the existence of fast degrees of freedom, like the vibrational degrees of freedom, to guaranty the existence of state with an internal temperature so that thermodynamic variables can be defined. This is clearly a physics inspired approximation.
We next define the internal clock which defines relaxation rate at the fictive temperature. The internal clock is different from the external clock.

Under nonisothermal conditions, the instantaneous relaxation rate changes with temperature; to preserve thermo‐rheological simplicity, one introduces the $\emph{material}$ or reduced time
\begin{equation}
\zeta(t)=\!\int_0^t\!\frac{dt'}{\tau[T(t'),T_f(t')]} .
\end{equation}
We next replace $t-s$ by $\zeta(t)-\zeta(s)$ in Eq.~(\ref{eq:ToolTime}) to obtain the generalized nonisothermal Tool–Narayanaswamy–Moynihan (TNM) expression,
\begin{equation}
T_f(t)
=T(t)-\!\int_0^t\!\Phi\!\big(\zeta(t)-\zeta(s)\big)\,\dot T(s)\,ds .
\label{eq:ToolReduced}
\end{equation}
Here $\Phi(\Delta\zeta)$ is the equilibrium structural‐relaxation function expressed in units of reduced time.  

This reparametrization makes the structural kernel depend only on the separation $\Delta\zeta$, implying that temperature modifies the rate but not the shape of relaxation (thermo‐rheological simplicity).

The reduced time $\zeta(t)$ accounts for internal relaxation time rather than laboratory seconds. this is consistent with the assumption of fictive temperature.

It advances rapidly when the structure can relax quickly (small $\tau$) and scarcely at all when relaxation is slow (large $\tau$). 
In effect, $\zeta$ acts as a \emph{relaxation odometer}: one unit of $\zeta$ corresponds to roughly one $e$-fold of structural decay. 
By writing the fictive temperature dynamics in reduced time, the theory ensures that $T_f(t)$ tracks the internal configurational state of the system regardless of the external cooling or heating schedule.

\section {Relation to  frequency dependent specific heat, $C_p''(\omega)$.}

As mentioned previously, one of the goals of this work is to relate the nonequilibrium relaxation function to frequency dependent specific heat by use of the established time correlation function.
Using 
\begin{equation}
\Phi(t)=\tfrac{2}{\pi}\!\int_0^\infty\![-C_p''(\omega)/(\Delta C_p\,\omega)]\cos(\omega t)\,d\omega, 
\end{equation}
Eq.~(\ref{eq:ToolReduced}) may be rewritten as
\begin{equation}
T_f(t)=T(t)
+\frac{2}{\pi}\!\int_0^\infty\!\frac{C_p''(\Omega)}{\Delta C_p\,\Omega}
\!\left[\int_0^t\!\cos\!\big(\Omega[\zeta(t)-\zeta(s)]\big)\,\dot T(s)\,ds\right] d\Omega,
\end{equation}
which expresses the fictive‐temperature evolution directly in terms of the experimentally measured calorimetric loss spectrum $C_p''(\omega)$.
This closes the bridge from microscopic enthalpy fluctuations to the macroscopic Tool equation used in nonisothermal glass dynamics.

Equations (7-8) use the \emph{equilibrium} kernel shape along an aging path via the reduced time $\zeta$ (quasi-equilibrium/material-time invariance). For strong driving, replace $\tau(T)\!\to\!\tau(T,T_f)$ in $\zeta$ (TNM) and, if needed, let the spectral weights evolve weakly with state.
%

%
\section{Use of VFT  for temperature dependence of viscosity \& the Maxwell viscoelastic relation}

We adopt a standard VFT law for viscosity (Pa\,s) with temperature in K:
\begin{equation}
\log_{10}\eta(T)=A+\frac{B}{T-T_0},\\
\qquad
A=-2.56,\;\; B=\SI{11520}{K},\;\; T_0=\SI{662}{K}.
\label{eq:VFT}
\end{equation}
The (Maxwell) structural relaxation time follows
\begin{equation}
\tau(T)=\frac{\eta(T)}{G_\infty},
\qquad
G_\infty\simeq \SI{31.4}{GPa}.
\end{equation}
VFT is operationally useful but diverges at $T\!\to\!T_0^+$; below $T_0$ the fit is not physically meaningful. In numerical implementation, we therefore \emph{freeze} the structure by capping $\tau$ to a large value when $T\le T_0$ (see Sec.~\ref{sec:numerics}).

\subsection{Determination of $\Phi$}
Two equivalent identification routes are standard.
\paragraph{(A) Frequency domain determination}
For small sinusoidal driving, the complex susceptibility (e.g.\ heat capacity) obeys
\begin{equation}
\frac{C^\ast(\omega)-C_\infty}{\Delta C}
=\int_{-\infty}^{\infty} \frac{G(\ln\hat\tau)}{1+i\omega\hat\tau}\,d\ln\hat\tau
=\mathcal{L}\{\Phi\}\big|_{s=i\omega}\!,
\end{equation}
so fitting $C^\ast(\omega)$ with a \emph{positive} spectrum.
determines $G$ and hence $\Phi$ via Eq.11.

\paragraph{(B) Time domain measurements.}
From a step or ramp experiment, invert  Laplace transformed form or fit a \emph{finite Prony spectrum} model which can be easily inverted analytically.
\begin{equation}
\Phi(\Delta\zeta)\approx\sum_{k=1}^{N} a_k\,e^{-\lambda_k\Delta\zeta},
\qquad a_k\ge 0,\ \sum a_k=1,\ \lambda_k>0,
\end{equation}
which can be embedded as $N$ auxiliary ODEs (stable, causal, Markovian).

\subsection{Commonly employed forms, and fits}
Debye (single mode)
\begin{equation}
\Phi_{\rm D}(\Delta\zeta)=e^{-\Delta\zeta},
\end{equation}
gives the ODE 
\begin{equation}
\dot T_f=\frac{T-T_f}{\tau(T)}.
\end{equation}
the stretched exponential form of KWW
\begin{equation}
\Phi_{\beta}(\Delta\zeta)=\exp[-(\Delta\zeta)^{\beta}],\ \ 0<\beta\le 1,
\end{equation}
gives broad spectrum.

The other popular form is Havriliak–Negami (HN):

\begin{equation}
\frac{C^\ast-C_\infty}{\Delta C}=\frac{1}{\left[1+(i\omega \tau_{\rm HN})^{\alpha}\right]^{\gamma}},
\end{equation}
Fit $(\alpha,\gamma)$; invert to $\Phi$ numerically.

\subsection{Requirement of self‐consistency}

In Tool–Narayanaswamy–Moynihan [2-3, 13-15] one usually holds the \emph{shape} $\Phi$ fixed and lets the
\emph{clock} be history dependent via $\tau(T,T_f)$ (self‐consistency). If needed, allow weak
shape drift, e.g.\ $\beta\!\to\!\beta(T,T_f)$ or $a_k\!\to\!a_k(T,T_f)$, but fit must preserve
positivity in $Phi$.
%
Equations (7) and (8) involve \emph{non-Markovian integral}: every past increment $\dot T(s)\,ds$ contributes to the present state with a weight 
\begin {equation}
M(t,s)=\Phi\big(\zeta(t)-\zeta(s)\big),\quad 0\le s\le t,
\end {equation}
that depends only on the reduced-time separation. Early times (small $\tau$) are quickly forgotten; near $T_g$ (huge $\tau$), reduced time accumulates slowly, so memory decays slowly in \emph{clock} time.

\subsection {Relation to the Debye single exponential description}
There are a few interesting consequences for the Debye relaxation process.

$\Phi(\Delta\zeta)=e^{-\Delta\zeta}$ is equivalent to the local ODE
\begin{equation}
\frac{dT_f}{dt} \;=\; \frac{T(t)-T_f(t)}{\tau\!\big(T(t)\big)} ,\qquad T_f(0)=T(0).
\label{eq:A5}
\end{equation}
This is the popular well-known equation for the fictive temperature.
Differentiate Eq.(6) to obtain

\begin{equation}
\frac{dT_f}{dt} = \dot T(t) - \int_{0}^{t} \frac{\partial}{\partial t}\Phi\!\big(\zeta(t)-\zeta(s)\big)\,\dot T(s)\,ds.
\end{equation}
Since $\partial_t\Phi(\zeta(t)-\zeta(s))=\Phi'(\Delta\zeta)\,\dot\zeta(t)= -e^{-\Delta\zeta}\,\frac{1}{\tau(T(t))}$, integration by parts recovers Eq.18. Conversely, solving the enthalpy equation with $T_f(0)=T(0)$ and substituting back reproduces Eq.18.

For a stretched response $\Phi(\Delta\zeta)=\exp[-(\Delta\zeta)^\beta]$ ($0<\beta<1$) or any positive Prony spectrum (which is a sum of damped exponentials)
\(\Phi(\Delta\zeta)=\sum_k a_k e^{-\lambda_k\Delta\zeta}\),
no single local ODE exists; the non-Markovian convolution integral. Alternatively, we can obtain  an equivalent set of embedded ODEs for the Prony modes) as the natural and numerically stable representation.

\section {Self‐consistency  in the TNM Model and frequency dependent specific heat}

This is more complicated.
To include structural feedback, replace \(\tau(T)\) by \(\tau(T,T_f)\) in the reduced-time clock,
\begin{equation}
\zeta_{TNM}(t)=\int_{0}^{t}\frac{dt'}{\tau\!\big(T(t'),T_f(t')\big)} ,
\label{eq:A6}
\end{equation}
and solve the TNM ordinary differential equation with Debye shape). 
This preserves the memory interpretation while making the clock history-dependent.

In the simplified Tool's integral equation, we used the $\zeta(t)$ defined earlier.

If $\Phi(\Delta\zeta)$ be the normalized structural response ($\Phi(0)=1$, $\Phi(\infty)=0$). The \emph{linear Tool} relation writes the fictive temperature as the Volterra convolution
\begin{equation}
T_f(t)
=
T(t)-\int_0^t \Phi\!\big(\zeta(t)-\zeta(s)\big)\,\frac{dT}{ds}(s)\,ds .
\label{eq:tool_general}
\end{equation}
For a linear cooling with rate q,  $dT/ds=-q$:
\begin{equation}
T_f(t)
=
T(t)+
q\int_0^t \Phi\!\big(\zeta(t)-\zeta(s)\big)\,ds .
\label{eq:tool_linear}
\end{equation}
Thus, if we know the frequency dependent specific heat, we can determine the fictive temperature from Eq.22.
We already mentioned common choices for the response functions.

\subsection {Ordinary differential equation (ODE) equivalence for Debye.}

With $\Phi_{\rm D}$, Eq.~\eqref{eq:tool_general} is equivalent to the local ODE
\begin{equation}
\frac{dT_f}{dt}=\frac{T(t)-T_f(t)}{\tau\!\big(T(t)\big)},\qquad T_f(0)=T(0).
\label{eq:tool_ode}
\end{equation}
%
%
\subsection{Assumption of thermorheological simplicity (TRS) and the form of $\Phi$ }

The key assumption we have made above is that of \emph{thermorheological simplicity} (TRS): the \emph{shape} of structural relaxation is temperature‐independent when time is measured in the \emph{reduced time} $\zeta$, i.e., after rescaling the physical time interval by the instantaneous relaxation time $\tau(T)$ [13-15].
Thus, the response is a function of $\Delta\zeta=\zeta(t)-\zeta(s)$ only, with
\begin{equation}
\Phi(0)=1,\qquad \lim_{\Delta\zeta\to\infty}\Phi(\Delta\zeta)=0,\qquad \Phi(\Delta\zeta)\ \text{is nonincreasing}.
\label{eq:phi_norm}
\end{equation}
{\it Physically, $\Phi$ is the fraction of the structural “departure from equilibrium’’ that remains after reduced time $\Delta\zeta$ has elapsed.}
\section{Nonlinear Tool--Narayanaswamy--Moynihan (TNM)}
We briefly discuss an extension of our discussion. This is the  common nonlinear generalization that uses a physics inspired combination of real temperature  $T$ and structurally determined fictive temperature ($T_f$) in the activation energy [13-15],
\begin{equation}
\ln \tau(T,T_f)=\ln \tau_0 + x\,\frac{E}{R\,T} + (1-x)\,\frac{E}{R\,T_f},\qquad 0\le x\le 1,
\label{eq:TNM_tau}
\end{equation}
with evolution
\begin{equation}
\frac{dT_f}{dt}=\frac{T(t)-T_f(t)}{\tau(T,T_f)} .
\label{eq:TNM_ode}
\end{equation}

\begin{equation}
\Phi_{\rm D}(\Delta\zeta)=e^{-\Delta\zeta},
\qquad
\Phi_{\beta}(\Delta\zeta)=\exp\!\big[-(\Delta\zeta)^{\beta}\big],\; 0<\beta\le 1 .
\tag{E8}
\end{equation}

\begin{equation}
\text{If }\Phi=\Phi_{\rm D}:\qquad
\frac{dT_f}{dt} \;=\; \frac{T(t)-T_f(t)}{\tau\!\big(T(t)\big)},
\qquad T_f(0)=T(0).
\tag{E9}
\end{equation}
%
\section {Self-consistent reduced time in TNM clock and evolution}
\addcontentsline{toc}{subsection}{Self-consistent reduced time (TNM clock) and evolution}

\begin{equation}
\ln \tau(T,T_f) \;=\; \ln \tau_0 \;+\; x\,\frac{E}{R\,T}
\;+\; (1-x)\,\frac{E}{R\,T_f},
\qquad 0\le x\le 1 .
\end{equation}

\begin{equation}
\zeta(t) \;=\; \int_0^{t} \frac{dt'}{\tau\!\big(T(t'),\,T_f(t')\big)} .
\end{equation}
%
\begin{equation}
\frac{dT_f}{dt} \;=\; \frac{T(t)-T_f(t)}{\tau\!\big(T(t),T_f(t)\big)} .
\end{equation}

\begin{equation}
T_f(t) \;=\; T(t)
\;-\; \int_0^t
\Phi\!\big(\zeta(t)-\zeta(s)\big)\,\frac{dT}{ds}(s)\,ds,
\quad
\zeta \text{ from \,(E11)} .
\end{equation}

\begin{equation}
\text{(Optional)}\quad \Phi \;\to\; \Phi\!\big(\zeta(t)-\zeta(s);\;\theta(s)\big),
\;\;\text{with } \theta \text{ a structure parameter (e.g.\ }\beta \text{ or Prony weights)} .
\end{equation}

%
\subsection{Implementing self-consistency in Tool-Narayanaswamy-Moynihan theory}
\addcontentsline{toc}{subsection}{Fixed-point self-consistency (for discussion)}

\begin{equation}
\text{Given } T(t),\ \text{ iterate:}\quad
\begin{cases}
\displaystyle \zeta^{(k+1)}(t) = \int_0^t \frac{dt'}{\tau\!\big(T(t'),\,T_f^{(k)}(t')\big)},\\[10pt]
\displaystyle T_f^{(k+1)}(t) = T(t)-\!\int_0^t \Phi\!\big(\zeta^{(k+1)}(t)-\zeta^{(k+1)}(s)\big)\,\dot T(s)\,ds,
\end{cases}
\ \ k=0,1,2,\dots
\end{equation}
with $T_f^{(0)}(t)=T(t)$ or $T_f^{(0)}(t)=T(0)$ a convenient initializer.
%
%
\section{Numerical analysis for silica glass}
\label{sec:numerics}
The relevant parameters of this problem are well-known, and well-documented in the literature, and can be given as follows.
\begin{equation}
T(t)=T_{\mathrm{start}}-q\,t,\qquad
T_{\mathrm{start}}=\SI{2000}{K},\quad
T_{\mathrm{end}}=\SI{500}{K},\quad
q=\SI{1}{K\,min^{-1}}.
\label{eq:cooling}
\end{equation}
The glass transition temperature of fused silica at ambient pressure is 662 K when its viscosity diverges.
\subsection{Temperature protocol, bath, and nonisothermal setup.}
We prescribe a temperature protocol $T(t)$ that the sample follows under strong thermal coupling to a programmable bath, so that external thermal equilibration is fast compared to structural relaxation. Under this assumption the fast vibrational enthalpy $H_{\infty}$ remains in instantaneous equilibrium with $T(t)$, while the slow configurational part $H_{c}$ lags behind and is represented by the fictive temperature $T_{f}(t)$. Because $T(t)$ varies during ramps or quenches, the calculation is \emph{nonisothermal}: the structural relaxation time $\tau$ changes along the path, so we measure history in the reduced (material) time $\zeta(t)=\int_{0}^{t} dt'/\tau(\cdot)$ and evolve $T_{f}$ using Tool or TNM kinetics. Unless stated otherwise, we assume ideal thermal contact; if necessary, a finite external thermal time $\tau_{\mathrm{ext}}=C_{\mathrm{tot}}/G_{\mathrm{th}}$ may be introduced via $\dot T=(T_{\mathrm{bath}}-T)/\tau_{\mathrm{ext}}$ with a prescribed bath program $T_{\mathrm{bath}}(t)$. In practice we specify: (i) the temperature protocol $T(t)$ (ramp rates and holds), (ii) the initial condition $T_{f}(0)$ (sample preparation), and (iii) the clock model $\tau$ (Maxwell–VFT or TNM $\tau(T,T_{f})$). We then integrate $T_{f}$ in time (or in reduced time) together with $\zeta(t)$, thereby quantifying the evolving internal dynamical state under arbitrary cooling or heating schedules.
\subsection{Choice of structural response function $\Phi$ in numerics.}
In the formal development above we expressed the structural kernel $\Phi(t)$ in terms of the
enthalpy–enthalpy time correlation function and, equivalently, the frequency-dependent
specific heat $C_p^\ast(\omega)$. In the numerical work below, however, we do not
compute $C_p^\ast(\omega)$ explicitly from a microscopic model (such as the Saito--Ohmine--Bagchi
molecular dynamics analysis of supercooled water\cite{SaitoOhmineBagchi2013}) but instead
\emph{assume} that the normalized structural response function in material time,
$\Phi(\Delta\zeta)$, is given by a Kohlrausch--Williams--Watts (KWW) form,
$\Phi(\Delta\zeta)=\exp[-(\Delta\zeta)^{\beta}]$ with $0<\beta\le1$. This may appear as a large
jump, but it is in fact a standard and well-justified phenomenological step. Experimental
dielectric, mechanical, and calorimetric spectra of supercooled liquids consistently show
broad, non-Debye relaxation, and the corresponding time-domain responses are accurately
described by stretched exponentials over many decades in time. A KWW ansatz is therefore a
compact way to encode a broad spectrum of relaxation times, in agreement with the Prony
representation implied by Bernstein's theorem for completely monotone kernels.
Within the present framework, any admissible $\Phi(\Delta\zeta)$ with a positive Prony spectrum
can be mapped to a unique complex heat capacity $C_p^\ast(\omega)$ and vice versa; choosing a
KWW shape simply means that we work with a \emph{model} $C_p^\ast(\omega)$ whose detailed
frequency dependence is not written explicitly but is implicitly defined as the Laplace
transform of the KWW kernel. [14,15] Our goal in the numerics is not to fit a specific experimental
$C_p^\ast(\omega)$ for silica but to explore how a broad, glass-like structural spectrum
influences the evolution of the fictive temperature $T_f(t)$ under different cooling protocols.
The Debye limit $\beta=1$ then recovers the simple Maxwell/Tool picture, while $0<\beta<1$
introduces the stretching and delayed relaxation that are ubiquitously observed in fragile and
strong glass formers alike. In this sense the use of a KWW structural response in $\zeta(t)$ is
a practical and physically motivated closure of the general fluctuation–response formalism
developed earlier.
%
\\
Silica is comparatively ``strong’’ liquid; near $T_g$ its structural relaxation is narrower than fragile
glasses. For first calculations use \textbf{Debye} (clarity, ODE form), then introduce
\textbf{KWW} with $0.6\lesssim\beta\lesssim 0.8$ (broader tail). For quantitative fitting,
use a \textbf{3–6 term Prony} spectrum [16,17] constrained by either $C^\ast(\omega)$ or a set of
DSC scans at multiple rates (global fit jointly with TNM clock parameters).


1) It cleanly separates the \emph{driving} ($\dot T$) from the \emph{structural memory} ($\Phi$), so arbitrary thermal histories (not just ramps) are straightforward.  
2) It works for \emph{nonexponential} relaxation (KWW), which is realistic for glasses.  
3) In the Debye limit it collapses to the simple ODE for $T_{f}(t)$ discussed above, letting us validate numerics by cross-checking both forms.

\subsection{ Inputs}

We briefly describe a few details of the numerical analysis employed to calculate time dependence of fictive temperature.

We take $T(t)$ as linear cooling with a fixed rate, VFT parameters from literature [13-15], $G_\infty = 30 GPa $ , and compute:

\begin{enumerate}
	\item $\eta(T)$ and $\tau(T)=\eta/G_\infty$ on a uniform time grid. To avoid the VFT singularity and keep the low-$T$ glass frozen, we impose
	\begin{equation}
	\tau(T)\;=\;\begin{cases}
	\eta(T)/G_\infty, & T>T_0,\\[2pt]
	\tau_{\max}, & T\le T_0,
	\end{cases}
	\qquad \tau_{\max}\gg \text{experiment time (here }10^{12}\,\mathrm{s}\text{)}.
	\label{eq:tau_cap}
	\end{equation}
	\item Reduced time by trapezoidal rule:
	\begin{equation}
	\zeta(t_i)\approx \sum_{k=1}^{i}\frac{1}{2}\Big(\frac{1}{\tau_{k}}+\frac{1}{\tau_{k-1}}\Big)\,\Delta t .
	\end{equation}
	\item \textbf{Tool (linear) Debye and KWW:} evaluate
	\begin{align}
	T_f^{(\mathrm{D})}(t_i) &= T(t_i)+q\int_0^{t_i}\!e^{-[\zeta(t_i)-\zeta(s)]}\,ds,\\
	T_f^{(\mathrm{KWW})}(t_i) &= T(t_i)+q\int_0^{t_i}\!\exp\!\big(-[\zeta(t_i)-\zeta(s)]^{\beta}\big)\,ds,
	\end{align}
	by trapezoidal quadrature on the time grid.
	%
	%
\end{enumerate}
We now present the numerical results obtained for silica by using the known parameters in Tool's equation, for
two different quenching rates. The figures illustrate how the fictive temperature behaves. Most importantly, it is seen to saturate, but in real world it would go to the final temperature but very slowly. That is why in specific heat heating experiments in cooling-heating cycle, the system is kept at the lowest temperature for a long time.
%
%
%
\begin{figure}[htbp]
	\centering
	\includegraphics[width=0.9\textwidth]{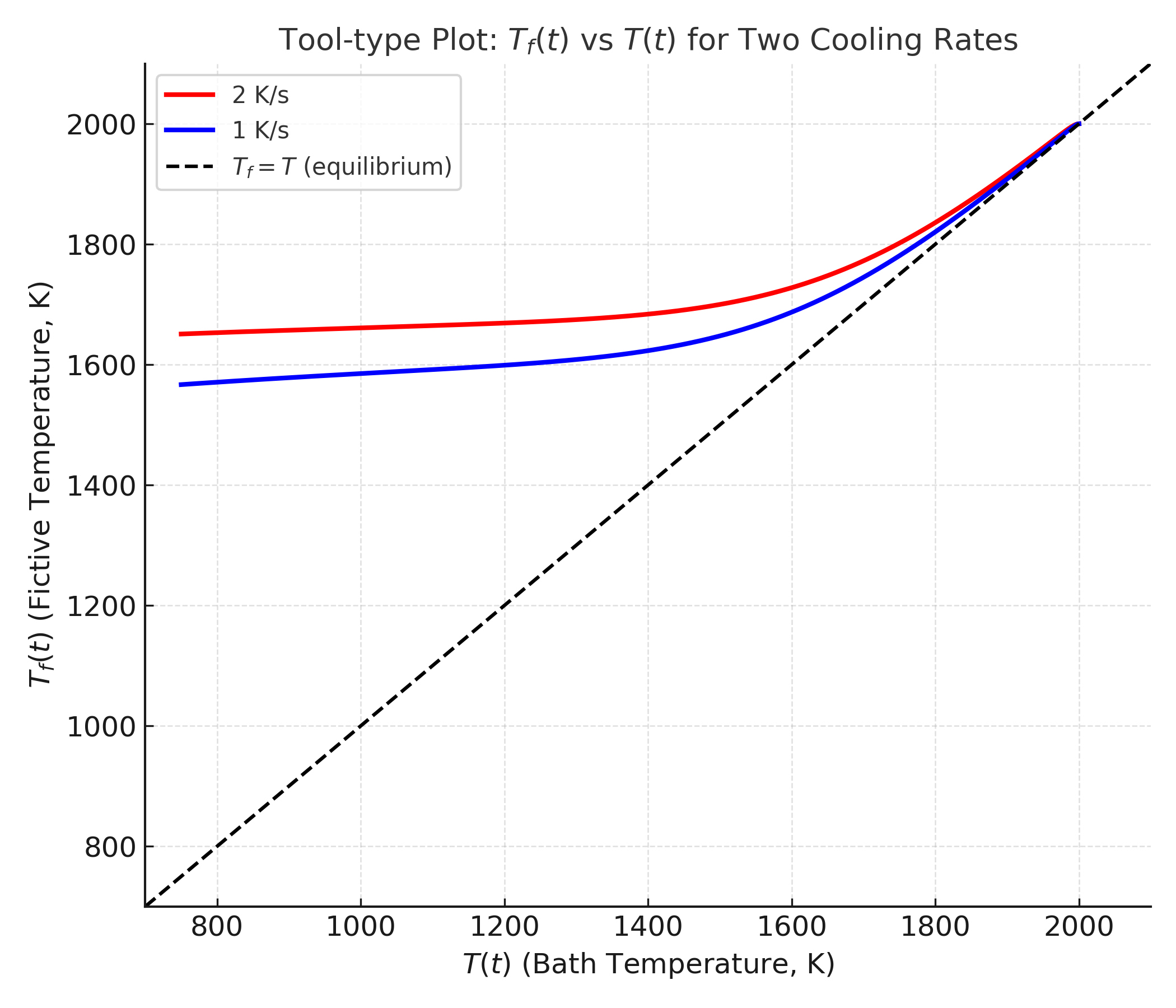}
	\label{fig:Fictive_Temp_Two_Rates}
	\caption{
		\textbf{Fictive temperature trajectories during linear cooling of silica.} The figure shows the Tool-type plots of the fictive temperature \(T_f(t)\) versus bath temperature \(T(t)\)
		for two different linear cooling rates: 1~K/s (blue) and 2~K/s (red), starting from 2000~K and ending at 750~K.
		The dashed black line denotes the equilibrium condition (\(T_f = T\)). At high temperatures, \(T_f\) follows \(T\) closely because structural relaxation is rapid and the system remains in equilibrium.
		As the temperature decreases through the glass-transition range (approximately 1300--1100~K),
		the structural relaxation time \(\tau(T)\) increases sharply, causing \(T_f\) to fall out of equilibrium
		and freeze at a constant value corresponding to the final structural state.
		The slower cooling rate (1~K/s) allows more time for structural relaxation,
		yielding a lower frozen-in fictive temperature compared to the faster 2~K/s quench. The curves were generated by numerically integrating the Tool equation 	$ \frac{dT_f}{dt} = \frac{T(t) - T_f(t)}{\tau(T)} $, using a logistic form for the structural relaxation time: $\tau(T) = \tau_{\infty} + \frac{\tau_0}{1 + \exp\!\left[\frac{T - T_g}{\Delta}\right]}$,
		where \(\tau_{\infty}=10^{-9}\,\mathrm{s}\), \(\tau_0=10^{4}\,\mathrm{s}\),
		\(T_g = 1200\,\mathrm{K}\), and \(\Delta = 100\,\mathrm{K}\).
		This choice ensures a smooth and physically realistic increase in \(\tau(T)\)from nanoseconds at high \(T\) to hours near the glass transition.}
 \label{fig:Fictive_Temp_Two_Rates}
	
\end{figure}

\end{document}